\documentclass[aps,prd,twocolumn,superscriptaddress]{revtex4}
\usepackage{epsfig,epsf}
\usepackage{amsmath}
\usepackage{amsthm}
\usepackage{amsfonts}
\usepackage{amssymb}
\usepackage{dsfont}
\usepackage{multirow}
\usepackage{appendix}
\usepackage{slashed}
\usepackage[active]{srcltx}
\usepackage{psfrag}

\begin{document}

\title{Fully heavy asymmetric scalar tetraquarks}
\date{\today}
\author{S.~S.~Agaev}
\affiliation{Institute for Physical Problems, Baku State University, Az--1148 Baku,
Azerbaijan}
\author{K.~Azizi}
\thanks{Corresponding Author}
\affiliation{Department of Physics, University of Tehran, North Karegar Avenue, Tehran
14395-547, Iran}
\affiliation{Department of Physics, Dogus University, Dudullu-\"{U}mraniye, 34775
Istanbul, T\"{u}rkiye}
\author{H.~Sundu}
\affiliation{Department of Physics Engineering, Istanbul Medeniyet University, 34700
Istanbul, T\"{u}rkiye}

\begin{abstract}
The scalar tetraquarks $T_{b}$ and $T_{c}$ with asymmetric contents $bb
\overline{b}\overline{c}$ and $cc \overline{c}\overline{b}$ are explored
using the QCD sum rule method. These states are modeled as the
diquark-antidiquarks composed of the axial-vector components. The masses and
current couplings of $T_{b}$ and $T_{c}$ are calculated using the two-point
sum rule approach. The predictions obtained for the masses of these
four-quark mesons prove that they are unstable against the strong two-meson
fall-apart decays to conventional mesons. In the case of the tetraquark $%
T_{b}$ this is the decay $T_{\mathrm{b}}\to \eta _{b}B_{c}^{-}$. The
processes $T_{\mathrm{c}}\rightarrow \eta _{c}B_{c}^{+}$ and $J/\psi
B_{c}^{\ast +}$ are kinematically allowed decay modes of the tetraquark $%
T_{c}$. The widths of corresponding processes are evaluated by employing the
QCD three-point sum rule approach which are necessary to estimate strong
couplings at the tetraquark-meson-meson vertices of interest. The mass $%
m=(15698 \pm 95)~\mathrm{MeV}$ and width $\Gamma[T_b]=(36.0 \pm 10.4)~
\mathrm{MeV}$ of the tetraquark $T_{b}$ as well as the parameters $%
\widetilde{m}=(9680 \pm 102)~\mathrm{MeV}$ and $\Gamma[T_c]=(54.7 \pm 12.6)~
\mathrm{MeV}$ in the case of $T_{c}$ provide useful information to search
for and interpret new exotic states.
\end{abstract}

\maketitle


\section{Introduction}

\label{sec:Intro}

Fully heavy tetraquarks, i.e., four-quark exotic mesons containing
exclusively heavy $b$ and $c$ quarks were and remain interesting objects for
theoretical investigations. Related problems were explored in numerous
publications aimed to calculate the masses of such hadrons, investigate
their stability against strong interactions and find kinematically allowed
decay channels. These studies permitted one to collect valuable information
about features of these hypothetical hadrons existence of which,
nevertheless, is not forbidden by laws of the parton model and quantum
choromodynamics. Researchers also elaborated new theoretical models and
methods to investigate these exotic hadrons.

Recently, the LHCb-ATLAS-CMS collaborations discovered four scalar $X$
resonances presumably with $cc\overline{c}\overline{c}$ contents \cite%
{LHCb:2020bwg,Bouhova-Thacker:2022vnt,CMS:2023owd}. This observation gave
strong impetus to physics of fully-heavy tetraquarks and placed it on the
strong foundation of experimental data. This achievement generated also new
theoretical works to interpret $X$ resonances as diquark-antidiquark states
or hadronic molecules, and explain their measured parameters (see, Refs.\
\cite{Agaev:2023wua,Agaev:2023ruu,Agaev:2023gaq,Agaev:2023rpj} and
references therein). In our articles, we computed the masses and widths of
the different models for the $X$ structures \cite%
{Agaev:2023wua,Agaev:2023ruu,Agaev:2023gaq,Agaev:2023rpj}. We found that
some of them may be considered as ground-level or radially excited
diquark-antidiquark states, whereas others probably are hadronic molecules
or superpositions of these two structures.

There are numerous fully heavy tetraquarks with interesting contents and
properties. The particles $bb\overline{c}\overline{c}$ are very intriguing
objects for studies, because they carry two units of electric charge. They
also may really be strong-interaction stable structures. The reason is that
their contents exclude creation of conventional mesons via the annihilation
of internal heavy quark-antiquark pairs originally studied in Refs.\ \cite%
{Becchi:2020mjz,Becchi:2020uvq}. Therefore, $bb\overline{c}\overline{c}$ may
be stable particles provided their masses are smaller than the corresponding
$B_{c}B_{c}$ meson thresholds. Our studies, however, demonstrated that the
masses of the tetraquarks $bb\overline{c}\overline{c}$ with quantum numbers $%
J^{\mathrm{P}}=0^{\pm },1^{\pm }$ and $2^{+}$ are above the relevant limits
and they are relatively wide compounds \cite%
{Agaev:2023tzi,Agaev:2024pej,Agaev:2024pil}.

Another class of fully heavy four-quark mesons is a family of hidden
charm-bottom particles $bc\overline{b}\overline{c}$. The structures $bc%
\overline{b}\overline{c}$ were not discovered yet, but have real chances to
be seen in ongoing and future experiments \cite%
{Carvalho:2015nqf,Abreu:2023wwg}. Properties of these tetraquarks were
studied in numerous publications by employing different methods and models
\cite%
{Faustov:2022mvs,Wu:2016vtq,Liu:2019zuc,Chen:2019vrj,Bedolla:2019zwg,Cordillo:2020sgc, Weng:2020jao,Yang:2021zrc,Hoffer:2024alv}%
. They were considered also in our articles \cite%
{Agaev:2024wvp,Agaev:2024mng,Agaev:2024qbh}, in which we calculated the
masses and widths of the scalar, axial-vector and tensor particles. Let us
note that the full widths of these tetraquarks were evaluated by taking into
account the fall-apart processes and decays triggered by $b\overline{b}$
annihilation inside of the exotic mesons. It turned out that these
tetraquarks are relatively wide structures with widths around of $100~%
\mathrm{MeV}$.

Apart from the tetraquarks discussed above there are particles with
asymmetric contents, i.e., particles composed of unequal number of $b$ and $%
c $ quarks $bb\overline{b}\overline{c}$ or $cc\overline{c}\overline{b}$.
They also attracted interests of researches and were explored in a number of
publications \cite%
{Liu:2019zuc,Deng:2020lqw,Yang:2021hrb,Mutuk:2022nkw,Zhang:2022qtp,Galkin:2023wox}%
, where the authors used various approaches ranging from a potential till
the relativistic quarks models. The predictions made in these articles differ from each other. The main parameter calculated there is the mass
of $bb\overline{b}\overline{c}$ or $cc\overline{c}\overline{b}$ states with
different spin-parities: Decays of these tetraquarks, as usual, were not
explored quantitatively. Therefore, there is a necessity to perform
comprehensive analysis of the asymmetric fully heavy four-quark systems by
including their decay modes.

In the current article, we investigate the scalar tetraquarks $bb\overline{b}%
\overline{c}$ or $cc\overline{c}\overline{b}$ and, in what follows, label
them as $T_{\mathrm{b}}$ and $T_{\mathrm{c}}$, respectively. We treat these
particles as the diquark-antidiquark compounds built of the axial-vector
diquark and antidiquark. We apply the two-point QCD sum rule method \cite%
{Shifman:1978bx,Shifman:1978by} to compute the masses and current couplings
of these structures. Our predictions for the masses of $T_{\mathrm{b}}$ and $%
T_{\mathrm{c}}$ demonstrate that they are unstable against strong
dissociations to two conventional mesons. In the case of $T_{\mathrm{b}}$
this is the decay $T_{\mathrm{b}}\rightarrow \eta _{b}B_{c}^{-}$. The
processes $T_{\mathrm{c}}\rightarrow \eta _{c}B_{c}^{+}$ and $J/\psi
B_{c}^{\ast +}$ are kinematically allowed fall-apart decays in the case of $%
T_{\mathrm{c}}$. To estimate partial widths of these modes, we employ the
three-point sum rule approach, which is required to evaluate the strong
couplings of tetraquarks $T_{\mathrm{b}}$ and $T_{\mathrm{c}}$ and final
conventional mesons.

This paper is structured in the following way: In Sec.\ \ref{sec:Mass}, we
compute the spectroscopic parameters of the scalar diquark-antidiquark
states $T_{\mathrm{b}}$ and $T_{\mathrm{c}}$. The widths of the decay $T_{%
\mathrm{b}}\rightarrow \eta _{b}B_{c}^{-}$ is calculated in Sec.\ \ref%
{sec:Widths1}. The full width of the tetraquark $T_{\mathrm{c}}$ saturated
by the channels $T_{\mathrm{c}}\rightarrow \eta _{c}B_{c}^{+}$ and $J/\psi
B_{c}^{\ast +}$ is evaluated in section \ref{sec:Widths2}. We make our
conclusions in the last part of the article, Sec. \ref{sec:Conc}.


\section{Spectroscopic parameters of the tetraquarks $T_{\mathrm{b}}$ and $%
T_{\mathrm{c}}$}

\label{sec:Mass}

Here, we concentrate on the masses and current couplings of the
diquark-antidiquark states $T_{\mathrm{b}}$ and $T_{\mathrm{c}}$ in the
context of QCD two-point sum rule approach. In this method one has to
introduce interpolating currents for the structures $T_{\mathrm{b}}$ and $T_{%
\mathrm{c}}$ and calculate relevant correlation functions. In its turn, the
interpolating currents are determined by the internal organizations of the
tetraquarks, i.e., by the spin-parities of the constituent diquarks and
antidiquarks. We are going to explain this question in the case of the
scalar structure $bb\overline{b}\overline{c}$: The same analysis is valid
for the exotic meson $cc\overline{c}\overline{b}$ as well.

In general, the scalar tetraquark $T_{\mathrm{b}}$ can be composed of five
different diquarks without derivatives \cite{Du:2012wp}. It is known that
structures built of the color antisymmetric diquarks, i.e., tetraquarks that
belong to $[\overline{\mathbf{3}}_{c}]\otimes \lbrack \mathbf{3}_{c}]$ or $[%
\mathbf{3}_{c}]\otimes \lbrack \overline{\mathbf{3}}_{c}]$ representations
of the color $SU_{c}(3)$ group are more favorable (have lower masses) than
ones containing color-symmetric diquarks \cite{Jaffe:2004ph}. But one should
take into account that the diquark $bb$ is the flavor-symmetric state.
Therefore, $T_{\mathrm{b}}$ has to be from $[\mathbf{6}_{f}]\otimes \lbrack
\overline{\mathbf{6}}_{f}]$ or $[\overline{\mathbf{6}}_{f}]\otimes \lbrack
\mathbf{6}_{f}]$ representations of the flavor group. Considering diquarks
with different spin-parities, it is not difficult to see that only diquarks $%
b^{T}C\gamma _{\mu }b$ and $b^{T}C\sigma _{\mu \nu }b$ belong to the $(%
\mathbf{6}_{f}\mathbf{,}\overline{\mathbf{3}}_{c})$ flavor-color
representation. As a result, corresponding antidiquarks $\overline{b}%
\overline{c}$ should be from $(\overline{\mathbf{6}}_{f}\mathbf{,3}_{c})$
representation which is necessary to form a flavor-color singlet state.

In general, it is possible to construct two structures $C\gamma _{\mu
}\otimes \gamma ^{\mu }C$ and $C\sigma _{\mu \nu }\otimes \sigma _{\mu \nu }C
$ with required features using diquarks $b^{T}C\gamma _{\mu }b$ and $%
b^{T}C\sigma _{\mu \nu }b$ and corresponding antidiquarks. We choose to
explore the tetraquark $T_{\mathrm{b}}$ composed of the axial-vector diquark
$b^{T}C\gamma _{\mu }b$ and antidiquark $\overline{b}\gamma ^{\mu }C%
\overline{c}$. As a result, we get the interpolating current $J(x)$

\begin{eqnarray}
J(x) &=&b_{a}^{T}(x)C\gamma _{\mu }b_{b}(x)\left[ \overline{b}_{a}(x)\gamma
^{\mu }C\overline{c}_{b}^{T}(x)\right.  \notag \\
&&\left. -\overline{b}_{b}(x)\gamma ^{\mu }C\overline{c}_{a}^{T}(x)\right] ,
\label{eq:CR1}
\end{eqnarray}%
where $C$ is the charge conjugation matrix, and $a$ and $b$ are the color
indices. This current describes the diquark-antidiquark state in $[\overline{%
\mathbf{3}}_{c}]\otimes \lbrack \mathbf{3}_{c}]$ representation of the color
group.

One of alternative options for $T_{\mathrm{b}}$ may be, for instance, a
structure built of the scalar diquark $b^{T}C\gamma _{5}b$ and antidiquark $%
\overline{b}\gamma _{5}C\overline{c}^{T}$. But $b^{T}C\gamma _{5}b$ is a
member of the representation $(\mathbf{6}_{f}\mathbf{,6}_{c})$, and such
current would correspond to the color-symmetric $[\mathbf{6}_{c}]\otimes
\lbrack \overline{\mathbf{6}}_{c}]$ tetraquark which is less stable than the
antisymmetric ones \cite{Jaffe:2004ph}.

The scalar particle $T_{\mathrm{c}}=cc\overline{c}\overline{b}$ is
constructed in analogous way and has the interpolating current
\begin{eqnarray}
\widetilde{J}(x) &=&c_{a}^{T}(x)C\gamma _{\mu }c_{b}(x)\left[ \overline{c}%
_{a}(x)\gamma ^{\mu }C\overline{b}_{b}^{T}(x)\right.  \notag \\
&&\left. -\overline{c}_{b}(x)\gamma ^{\mu }C\overline{b}_{a}^{T}(x)\right] .
\label{eq:CR2}
\end{eqnarray}


\subsection{The mass and current coupling of the scalar state $bb\overline{b}%
\overline{c}$}


We are going to derive the expressions for the mass $m$ and current coupling
$\Lambda $ of the tetraquark $T_{\mathrm{b}}$ in the context of the QCD sum
rule method \cite{Shifman:1978bx,Shifman:1978by}. It is one of the powerful
nonperturbative methods to extract parameters of the hadrons. Originally, it
was proposed to study conventional particles, but is successfully used also
to explore multiquark systems \cite{Albuquerque:2018jkn,Agaev:2020zad}.

We start our investigations from the correlation function
\begin{equation}
\Pi (p)=i\int d^{4}xe^{ipx}\langle 0|\mathcal{T}\{J(x)J^{\dag
}(0)\}|0\rangle ,  \label{eq:CF1}
\end{equation}%
where $\mathcal{T}$ \ is the time-ordered product of two currents. In
accordance with SR method, the correlation function $\Pi (p)$ should be
expressed by utilizing the physical parameters of the tetraquark $m$ and $%
\Lambda $, and computed in the operator product expansion ($\mathrm{OPE}$)
with fixed accuracy by employing heavy quark propagators. The first
expression forms the phenomenological side $\Pi ^{\mathrm{Phys.}}(p)$ of the
SR equality, whereas the second one $\Pi ^{\mathrm{OPE}}(p)$--its QCD side.
Afterwards, by matching these two expressions within the hadron-quark
duality assumption, and carrying out the manipulations explained below, one
finds required SRs.

The correlator $\Pi ^{\mathrm{Phys}}(p)$ is given by the expression
\begin{equation}
\Pi ^{\mathrm{Phys}}(p)=\frac{\langle 0|J|T_{\mathrm{b}}\rangle \langle T_{%
\mathrm{b}}|J^{\dagger }|0\rangle }{m^{2}-p^{2}}+\cdots ,  \label{eq:Phys1}
\end{equation}%
which is the sum of the ground-level particle's contribution shown
explicitly, and higher resonances and continuum states denoted by the
ellipses.

We rewrite $\Pi ^{\mathrm{Phys}}(p)$ by utilizing the matrix element
\begin{equation}
\langle 0|J|T_{\mathrm{b}}\rangle =\Lambda ,  \label{eq:ME1}
\end{equation}%
and get
\begin{equation}
\Pi ^{\mathrm{Phys}}(p)=\frac{\Lambda ^{2}}{m^{2}-p^{2}}+\cdots .
\label{eq:Phys2}
\end{equation}%
The term $\Lambda ^{2}/(m^{2}-p^{2})$ in the right-hand side of Eq.\ (\ref%
{eq:Phys2}) has the trivial Lorentz structure proportional to $\mathrm{I}$,
and is the invariant amplitude $\Pi ^{\mathrm{Phys}}(p^{2})$ required for
future studies.

Alternatively, $\Pi ^{\mathrm{Phys.}}(p^{2})$ can be expressed\ through
dispersion integral in terms of the spectral density $\rho ^{\mathrm{Phys}%
}(s)$
\begin{equation}
\rho ^{\mathrm{Phys}}(s)=\Lambda ^{2}\delta (s-m^{2})+\rho ^{\mathrm{h}%
}(s)\theta (s-s_{0}),  \label{eq:SDensity}
\end{equation}%
where $\theta (s-s_{0})$ is the unit step function, and $s_{0}$ is the
continuum subtraction parameter. Here, the $\Lambda ^{2}\delta (s-m^{2})$ is
the pole term, whereas $\rho ^{\mathrm{h}}(s)$ is unknown hadronic spectral
density. Then, it is evident that
\begin{equation}
\Pi ^{\mathrm{Phys}}(p^{2})=\frac{\Lambda ^{2}}{m^{2}-p^{2}}%
+\int_{s_{0}}^{\infty }\frac{\rho ^{\mathrm{h}}(s)ds}{s-p^{2}}.
\label{eq:InvAmp2}
\end{equation}%
In the region $p^{2}\ll 0$ we apply the Borel transformation $\mathcal{B}$
to remove subtraction terms in the dispersion integral and suppress
contributions of higher resonances and continuum states. As a result, for $%
\mathcal{B}\Pi ^{\mathrm{Phys}}(p^{2})$ we obtain%
\begin{equation}
\mathcal{B}\Pi ^{\mathrm{Phys}}(p^{2})=\Lambda
^{2}e^{-m^{2}/M^{2}}+\int_{s_{0}}^{\infty }ds\rho ^{\mathrm{h}%
}(s)e^{-s/M^{2}},  \label{eq:CorBor}
\end{equation}%
where $M^{2}$ is the Borel parameter.

At next phase of studies, we insert the current $J(x)$ into Eq.\ (\ref%
{eq:CF1}) and contract relevant quark fields to find $\Pi ^{\mathrm{OPE}}(p)$%
, which reads%
\begin{eqnarray}
&&\Pi ^{\mathrm{OPE}}(p)=2i\int d^{4}xe^{ipx}\mathrm{Tr}\left[
S_{b}^{b^{\prime }a}(-x)\gamma _{\mu }\widetilde{S}_{c}^{a^{\prime
}b}(-x)\gamma _{\nu }\right]  \notag \\
&&\times \left\{ \mathrm{Tr}\left[ \gamma ^{\nu }\widetilde{S}%
_{b}^{aa^{\prime }}(x)\gamma ^{\mu }S_{b}^{bb^{\prime }}(x)\right] +\mathrm{%
Tr}\left[ \gamma ^{\nu }\widetilde{S}_{b}^{bb^{\prime }}(x)\gamma ^{\mu
}S_{b}^{aa^{\prime }}(x)\right] \right.  \notag \\
&&\left. -\mathrm{Tr}\left[ \gamma ^{\nu }\widetilde{S}_{b}^{ba^{\prime
}}(x)\gamma ^{\mu }S_{b}^{ab^{\prime }}(x)\right] -\mathrm{Tr}\left[ \gamma
^{\nu }\widetilde{S}_{b}^{ab^{\prime }}(x)\gamma ^{\mu }S_{b}^{ba^{\prime
}}(x)\right] \right\} ,  \notag \\
&&  \label{eq:QCD1}
\end{eqnarray}%
where
\begin{equation}
\widetilde{S}_{b(c)}(x)=CS_{b(c)}^{T}(x)C.  \label{eq:Prop}
\end{equation}%
Here, $S_{b(c)}(x)$ are $b$ and $c$ quarks' propagators \cite{Agaev:2020zad}.

The amplitude $\Pi ^{\mathrm{OPE}}(p^{2})$ is calculated in deep Euclidean
region $p^{2}\ll 0$ where coefficient functions in $\mathrm{OPE}$ is
obtained using the perturbative QCD, whereas nonperturbative information is
contained in the gluon condensates. Having continued $\Pi ^{\mathrm{OPE}%
}(p^{2})$ analytically to the Minkowski domain and found its imaginary part,
we get the two-point spectral density $\rho ^{\mathrm{OPE}}(s)$. One can
write the dispersion representation for the amplitude $\Pi ^{\mathrm{OPE}%
}(p^{2})$ using $\rho ^{\mathrm{OPE}}(s)$ in the range $s\in \lbrack
\mathcal{M}^{2},\infty ]$ where $\mathcal{M}=3m_{b}+m_{c}$. Then, by
equating the Borel transformations of $\Pi ^{\mathrm{Phys}}(p^{2})$ and $\Pi
^{\mathrm{OPE}}(p^{2})$ and applying the assumption about hadron-parton
duality $\rho ^{\mathrm{h}}(s)\simeq \rho ^{\mathrm{OPE}}(s)$ in a duality
region, we subtract second term in Eq.\ (\ref{eq:CorBor}) from the QCD side
of the obtained equality and get
\begin{equation}
\Lambda ^{2}e^{-m^{2}/M^{2}}=\Pi (M^{2},s_{0}).  \label{eq:SR}
\end{equation}%
Here,
\begin{equation}
\Pi (M^{2},s_{0})=\int_{\mathcal{M}^{2}}^{s_{0}}ds\rho ^{\mathrm{OPE}%
}(s)e^{-s/M^{2}}+\Pi (M^{2}).  \label{eq:CorrF}
\end{equation}%
The nonperturbative function $\Pi (M^{2})$ is computed directly from the
correlator $\Pi ^{\mathrm{OPE}}(p)$ and contains contributions that do not
enter to the spectral density.

After simple manipulations, we get
\begin{equation}
m^{2}=\frac{\Pi ^{\prime }(M^{2},s_{0})}{\Pi (M^{2},s_{0})},  \label{eq:Mass}
\end{equation}%
and
\begin{equation}
\Lambda ^{2}=e^{m^{2}/M^{2}}\Pi (M^{2},s_{0}),  \label{eq:Coupl}
\end{equation}%
which are the sum rules for $m$ and $\Lambda $, respectively. In Eq.\ (\ref%
{eq:Mass}), we also use $\Pi ^{\prime }(M^{2},s_{0})=d\Pi
(M^{2},s_{0})/d(-1/M^{2})$. The spectral density $\rho ^{\mathrm{OPE}}(s)$
contains the perturbative $\rho ^{\mathrm{pert.}}(s)$ and nonperturbative $%
\rho ^{\mathrm{Dim4}}(s)$ terms. We do not provide the explicit formulas for
the functions $\rho ^{\mathrm{OPE}}(s)$ and $\Pi (M^{2})$, because they are
rather lengthy.

We need to specify the input parameters in Eqs.\ (\ref{eq:Mass}) and (\ref%
{eq:Coupl}) to perform numerical computations. Some of them are universal
quantities: The masses of $b$ and $c$ quarks and gluon vacuum condensate $%
\langle \alpha _{s}G^{2}/\pi \rangle $ are such parameters. In the present
work, we use following values
\begin{eqnarray}
&&m_{b}(\mu )=4.18_{-0.02}^{+0.03}~\mathrm{GeV},\ m_{c}(\mu )=(1.27\pm 0.02)~%
\mathrm{GeV},  \notag \\
&&\langle \alpha _{s}G^{2}/\pi \rangle =(0.012\pm 0.004)~\mathrm{GeV}^{4}.
\label{eq:GluonCond}
\end{eqnarray}%
The $m_{b}(\mu )$ and $m_{c}(\mu )$ are the running quark masses in the $%
\overline{\mathrm{MS}}$ scheme at scales $\mu =m_{b}$ and $\mu =m_{c}$ \cite%
{PDG:2024}, respectively. The gluon vacuum condensate was extracted from
analysis of various hadronic processes in Refs.\ \cite%
{Shifman:1978bx,Shifman:1978by}.

Contrary, the Borel and continuum subtraction parameters $M^{2}$ and $s_{0}$
are specific for each problem and should satisfy some standard constraints
of SR computations. Dominance of the pole contribution ($\mathrm{PC}$) in
extracted quantities and their stability upon variations of $M^{2}$ and $%
s_{0}$ as well as convergence of the operator product expansion are
important conditions for correct SR analysis. To fulfill these requirements,
we impose on the parameters $M^{2}$ and $s_{0}$ the following restrictions.
First, the pole contribution
\begin{equation}
\mathrm{PC}=\frac{\Pi (M^{2},s_{0})}{\Pi (M^{2},\infty )},  \label{eq:PC}
\end{equation}%
should obey $\mathrm{PC}\geq 0.5$. The convergence of $\mathrm{OPE}$ is
second important condition in the SR analysis. Because, the correlation
function contains only nonperturbative dimension-$4$ term $\Pi ^{\mathrm{Dim4%
}}(M^{2},s_{0})$, we require fulfilment of the constraint $|\Pi ^{\mathrm{%
Dim4}}(M^{2},s_{0})|=0.05\Pi (M^{2},s_{0})$, which ensures the convergence
of the operator product expansion. It is worth noting that the maximum of
the Borel parameter is determined from Eq.\ (\ref{eq:PC}), whereas
convergence of $\mathrm{OPE}$ allows us to fix its minimal value.

\begin{figure}[h]
\includegraphics[width=8.5cm]{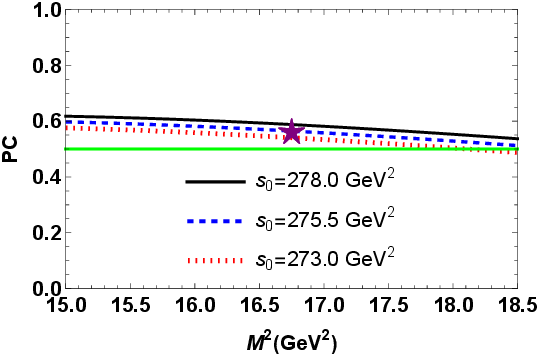}
\caption{Dependence of $\mathrm{PC}$ on the Borel parameter $M^{2}$ at fixed
$s_{0}$. The red star shows the point $M^{2}=16.75~\mathrm{GeV}^{2}$ and $%
s_{0}=275.5~\mathrm{GeV}^{2}$. }
\label{fig:PC}
\end{figure}

Numerical calculations are performed over a wide range of the parameters $%
M^{2}$ and $s_{0}$. Analysis of these predictions allows us to fix the
working windows for $M^{2}$ and $s_{0}$, where all aforementioned
restrictions are obeyed. We find that the regions
\begin{equation}
M^{2}\in \lbrack 15,18.5]~\mathrm{GeV}^{2},\ s_{0}\in \lbrack 273,278]~%
\mathrm{GeV}^{2},  \label{eq:Wind1}
\end{equation}%
comply with these constraints. Indeed, on the average in $s_{0}$ at maximal
and minimal $M^{2}$ the pole contribution is $\mathrm{PC}\approx 0.51$ and $%
\mathrm{PC}\approx 0.6$, respectively. The nonperturbative term is positive
and at $M^{2}=15~\mathrm{GeV}^{2}$ forms less than $1\%$ of the whole
result. The dependence of $\mathrm{PC}$ on the Borel parameter is shown in
Fig.\ \ref{fig:PC}, in which all curves exceed the limit line $\mathrm{PC}%
=0.5$.

To extract $m$ and $\Lambda $, we compute their mean values over the regions
Eq.\ (\ref{eq:Wind1}) and find
\begin{eqnarray}
m &=&(15698\pm 95)~\mathrm{MeV},  \notag \\
\Lambda &=&(9.30\pm 1.03)~\mathrm{GeV}^{5}.  \label{eq:Result1}
\end{eqnarray}%
Effectively, results in Eq.\ (\ref{eq:Result1}) are equal to SR predictions
at the point $M^{2}=16.75~\mathrm{GeV}^{2}$ and $s_{0}=275.5~\mathrm{GeV}%
^{2} $, where $\mathrm{PC}\approx 0.56$, which ensures the dominance of $%
\mathrm{PC}$ in the extracted parameters. The uncertainties in Eq.\ (\ref%
{eq:Result1}) are generated mainly by the choices of $M^{2}$ and $s_{0}$:
Ambiguities connected with variations of the quark masses $m_{b}$ and $m_{c}$
are small. Modifications due to uncertainties in the gluon condensate are
negligible and can be safely neglected. A scale dependence of these
parameters is also a possible source of ambiguities. But the gluon
condensate $\langle \alpha _{s}G^{2}/\pi \rangle $ is the $\mu $-scale
independent parameter, whereas rescaling in $m_{b}(\mu )$ and $m_{c}(\mu )$
can be treated as their uncertainties which have relatively small impact on
extracted quantities. Therefore, throughout present analysis for $m_{b}$ and
$m_{c}$ we employ the $\overline{\mathrm{MS}}$ results at fixed scales.

Note that the theoretical errors form only $\pm 0.6\%$ of the mass $m$,
which demonstrates the high stability of the obtained prediction. Such
accuracy of the result is connected with the correctness of the working
regions for $M^{2}$ and $s_{0}$ and analytical form of the SR for $m^{2}$ in
Eq.\ (\ref{eq:Mass}). Indeed, the SR determines $m$ as a ratio of the
correlation functions. Therefore, changes in the correlators due to $M^{2}$,
$s_{0}$ and another parameters compensate in $m$ each other and stabilize by
this way the numerical output. In the case of $\Lambda $ errors amount to $%
\pm 11\%$ of the central value, but still remain within limits acceptable of
the sum rule analysis. In Fig.\ \ref{fig:Mass}, we show $m$ as a function of
$M^{2}$ and $s_{0}$.

\begin{widetext}

\begin{figure}[h!]
\begin{center}
\includegraphics[totalheight=6cm,width=8cm]{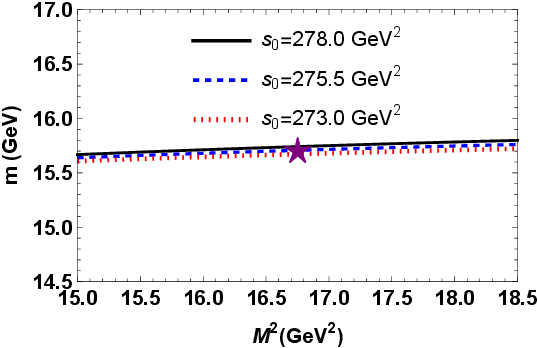}
\includegraphics[totalheight=6cm,width=8cm]{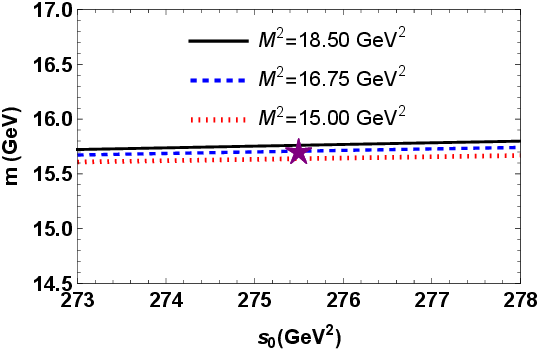}
\end{center}
\caption{The mass $m$ as a function of the Borel  $M^{2}$ (left panel), and continuum threshold $s_0$ parameters (right panel).}
\label{fig:Mass}
\end{figure}

\begin{figure}[h!]
\begin{center}
\includegraphics[totalheight=6cm,width=8cm]{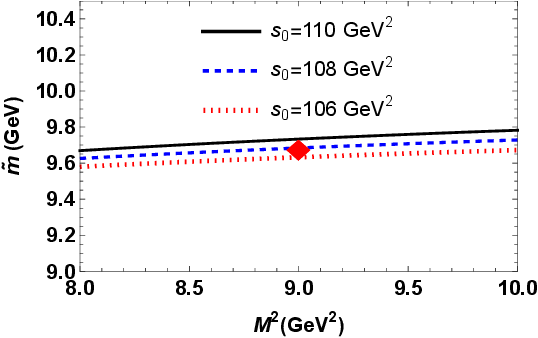}
\includegraphics[totalheight=6cm,width=8cm]{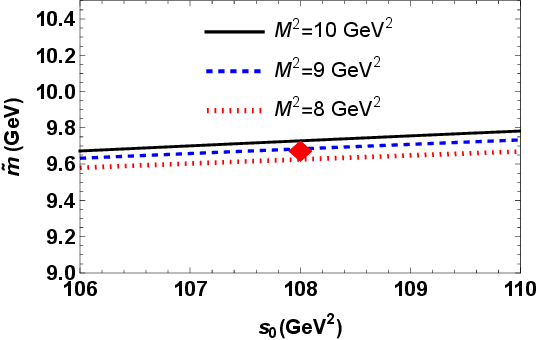}
\end{center}
\caption{Dependence of $\widetilde{m}$ on the Borel  $M^{2}$ (left panel), and continuum threshold $s_0$ parameters (right panel). The red diamond is placed at $M^{2}=9~\mathrm{GeV}^{2}$ and
$s_{0}=108~\mathrm{GeV}^{2}$.}
\label{fig:Mass1}
\end{figure}

\end{widetext}


\subsection{Parameters of the tetraquark $T_{\mathrm{c}}$}


The expressions for the correlation functions $\widetilde{\Pi }^{\mathrm{Phys%
}}(p)$ and $\widetilde{\Pi }^{\mathrm{OPE}}(p)$, as well as the SRs for $%
\widetilde{m}$ and $\widetilde{\Lambda }$ in the case of the tetraquark $T_{%
\mathrm{c}}$ with content $cc\overline{c}\overline{b}$ differ from ones
obtained in the subsection by replacements $m_{b}\leftrightarrow m_{c}$.
Therefore, we omit related details and write down only the working intervals
for the auxiliary parameters $M^{2}$ and $s_{0}$. Numerical computations
prove that
\begin{equation}
M^{2}\in \lbrack 8,10]~\mathrm{GeV}^{2},\ s_{0}\in \lbrack 106,110]~\mathrm{%
GeV}^{2},  \label{eq:Wind1A}
\end{equation}%
comply with all necessary constraints. Really, on the average in $s_{0}$ at
maximal $M^{2}=10~\mathrm{GeV}^{2}$ the pole contribution is $\mathrm{PC}%
\approx 0.50$, whereas at $M^{2}=8~\mathrm{GeV}^{2}$ it is equal to $\mathrm{%
PC}\approx 0.73$. The nonperturbative term is positive and at $M^{2}=8~%
\mathrm{GeV}^{2}$ forms $0.4\%$ of the whole result.

To extract $\widetilde{m}$ and $\widetilde{\Lambda }$, we compute their
average values over the regions Eq.\ (\ref{eq:Wind1A}) and find
\begin{eqnarray}
\widetilde{m} &=&(9680\pm 102)~\mathrm{MeV},  \notag \\
\widetilde{\Lambda } &=&(1.55\pm 0.22)~\mathrm{GeV}^{5}.  \label{eq:Result2}
\end{eqnarray}%
Dependence of the mass $\widetilde{m}$ on the Borel parameter $M^{2}$ and $%
s_{0}$ is depicted in Fig.\ \ref{fig:Mass1}.

The diquark-antidiquark states $bb\overline{b}\overline{c}$ and $cc\overline{%
c}\overline{b}$ with different spin-parities were investigated in various
articles \cite%
{Liu:2019zuc,Deng:2020lqw,Yang:2021hrb,Mutuk:2022nkw,Zhang:2022qtp,Galkin:2023wox}%
. It is interesting to compare our findings with predictions for the mass of
the scalar tetraquarks made in these works. Thus, in Ref.\ \cite{Liu:2019zuc}
the authors used the nonrelativistic quark model where the Hamiltonian
contains the linear confining and Coulomb potentials and spin-spin
interactions. The results for the mass of the tetraquarks $bb\overline{b}%
\overline{c}$ composed of the color-triplet and color-sextet diquarks were
found equal to $16158~\mathrm{MeV}$ and $16173~\mathrm{MeV}$, respectively.
In this paper, the mass of the scalar states $cc\overline{c}\overline{b}$
were estimated $9740~\mathrm{MeV}$ and $9763~\mathrm{MeV}$.

Three different models, i.e., the color-magnetic interaction model, the
constituent quark and multiquark flux-tube models were used in Ref.\ \cite%
{Deng:2020lqw} to evaluate the mass of the tetraquarks under consideration.
The authors considered the superpositions of the states built of
color-triplet and -sextet diquarks. For the mass of the scalar state $bb%
\overline{b}\overline{c}$ the color-magnetic interaction model predicted $%
15713~\mathrm{MeV}$, whereas the constituent quark model led to the result $%
16175~\mathrm{MeV}$. As is seen, there is an approximately $300~\mathrm{MeV}$
mass gap between two models. The same is true in the case of the tetraquark $%
cc\overline{c}\overline{b\text{ }}$ as well: $9314~\mathrm{MeV}$ and $9753~%
\mathrm{MeV}$, respectively.

The dynamical diquark and the relativistic diquark-antidiquark models \cite%
{Mutuk:2022nkw,Galkin:2023wox} gave $16060~\mathrm{MeV}$ and $16102~\mathrm{%
MeV,}$ respectively. For the tetraquark $cc\overline{c}\overline{b}$ the
authors found $9579~\mathrm{MeV}$ and $9606~\mathrm{MeV}$. Let us note that
these predictions corresponds to pure $[\overline{\mathbf{3}}_{c}]\otimes
\lbrack \mathbf{3}_{c}]$ diquark-antidiquark states.

Results of model analyses depend on diquark-antidiquark and/or
quark-antiquark interactions taken into account in relevant Hamiltonians, on
nonrelativistic or relativistic treatment of these interactions, on a
difference in constituent quark masses used in computations. There are big
gaps between masses of the same particles obtained in the framework of
different models \cite{Deng:2020lqw}, as it has just been emphasized above.
Stated differently, there is not consensus among different quark models on
the masses of the tetraquarks $T_{\mathrm{b}}$ and $T_{\mathrm{c}}$. It is
also difficult to trace the cause of these discrepancies because an each
model addresses some limited type of interactions in the diquark-antidiquark
systems.

The sum rule approach relies on first principles of the quark-gluon
interactions and uses quantum field-theoretical methods in corresponding
analysis. This analysis employs a few well-known universal parameters such
as masses of the quarks and gluon condensate(s). The only free quantities in
the SR computations are the Borel $M^{2}$ and continuum subtraction $s_{0}$
parameters. But their choice is restricted by strong conditions of the SR
method, therefore $M^{2}$ and $s_{0}$ can be varied within fixed limits. At
the same time, variations of $M^{2}$ and $s_{0}$ enable us to estimate
accuracy of obtained results, which is almost absent in quark-model studies.

Comparing our SR predictions $m=15698~\mathrm{MeV}$ and $\widetilde{m}=9680~%
\mathrm{MeV}$ with ones obtained in the context of the different quark
models, we see that the mass of the $T_{\mathrm{b}}$ state is compatible
with prediction of the color-magnetic interaction model \cite{Deng:2020lqw}
and is considerably smaller than results of other works. Contrary, the mass $%
\widetilde{m}$ of the tetraquark $T_{\mathrm{c}}$, in some cases, exceeds
predictions of the alternative approaches. It is interesting to note that
the difference $(3m_{b}+m_{c})-(m_{b}+3m_{c})=5820~\mathrm{MeV}$ is close to
our finding $m-\widetilde{m}\approx 6000~\mathrm{MeV}$. Corresponding mass
gaps in the quark models is about $300-400~\mathrm{MeV}$ larger than this
value.


\section{Decay $T_{\mathrm{b}}\rightarrow \protect\eta _{b}B_{c}^{-}$}

\label{sec:Widths1}

Information on the mass of the scalar state $T_{\mathrm{b}}$ allows us to
fix its decay channels. The decay to $\eta _{b}B_{c}^{-}$ mesons is
kinematically possible process for the transformation of the tetraquark $T_{%
\mathrm{b}}$ to ordinary mesons. The mesons $\eta _{b}$ and $B_{c}^{-}$ have
the masses $m_{\eta _{b}}=(9398.7\pm 2.0)~\mathrm{MeV}$ and $%
m_{B_{c}}=(6274.47\pm 0.27\pm 0.17)~\mathrm{MeV}$ \cite{PDG:2024},
respectively. As a result, the threshold for creation of the final state $%
\eta _{b}B_{c}^{-}$ is $15673~\mathrm{MeV}$. The central value $15698~%
\mathrm{MeV}$ for the mass of the tetraquark $T_{\mathrm{b}}$ is only $25~%
\mathrm{MeV}$ above this threshold. In the low limit of $m$, i.e., at $15603~%
\mathrm{MeV}$ the state $T_{\mathrm{b}}$ becomes stable against such
fall-apart processes. Therefore, in our analysis we will calculate the width
of $T_{\mathrm{b}}$ using the fixed mass $15698~\mathrm{MeV}$.

The width of the decay $T_{\mathrm{b}}(p)\rightarrow \eta _{b}(p^{\prime
})B_{c}^{-}(q)$, besides the usual input parameters is determined by the
strong coupling $G$ at the vertex $T_{\mathrm{b}}\rightarrow \eta
_{b}B_{c}^{-}$. It can be calculated by employing the form factor $G(q^{2})$
at the mass shell $q^{2}=m_{B_{c}}^{2}$. In its turn, the form factor $%
G(q^{2})$ can be extracted from analysis of the correlation function
\begin{eqnarray}
\Pi (p,p^{\prime }) &=&i^{2}\int d^{4}xd^{4}ye^{ip^{\prime
}y}e^{-ipx}\langle 0|\mathcal{T}\{J^{\eta _{b}}(y)  \notag \\
&&\times J^{B_{c}^{-}}(0)J^{\dagger }(x)\}|0\rangle ,  \label{eq:CF1a}
\end{eqnarray}%
where $J^{\eta _{b}}(x)$ and $J^{B_{c}^{-}}(x)$ are interpolating currents
of the pseudoscalar mesons $\eta _{b}$ and $B_{c}^{-}$, respectively
\begin{equation}
J^{\eta _{b}}(x)=\overline{b}_{i}(x)i\gamma _{5}b_{i}(x),\ J^{B_{c}^{-}}(x)=%
\overline{c}_{j}(x)i\gamma _{5}b_{j}(x).
\end{equation}%
Here, $i$ and $j$ are the color indices. The four-momenta of particles $p$, $%
p^{\prime }$ and $q$ are related by the formula $p=p^{\prime }+q$.

To find the physical side of the sum rule $\Pi ^{\mathrm{Phys}}(p,p^{\prime
})$, we have to write Eq.\ (\ref{eq:CF1a}) using parameters of the particles
$T_{\mathrm{b}}$, $\eta _{b}$ and $B_{c}^{-}$. To this end, we insert
complete sets of intermediate states for the particles $T_{\mathrm{b}}$, $%
\eta _{b}$ and $B_{c}^{-}$ into Eq.\ (\ref{eq:CF1a}) and perform
four-integrals over $x$ and $y$. After separating the contribution of the
ground-level particles together with a naive factorization approximation, we
get
\begin{eqnarray}
&&\Pi ^{\mathrm{Phys}}(p,p^{\prime })=\frac{\langle 0|J^{\eta _{b}}|\eta
_{b}(p^{\prime })\rangle }{p^{\prime 2}-m_{\eta _{b}}^{2}}\frac{\langle
0|J^{B_{c}^{-}}|B_{c}^{-}(q)\rangle }{q^{2}-m_{B_{c}}^{2}}  \notag \\
&&\times \langle \eta _{b}(p^{\prime })B_{c}^{-}(q)|T_{\mathrm{b}}(p)\rangle
\frac{\langle T_{\mathrm{b}}(p)|J^{\dagger }|0\rangle }{p^{2}-m^{2}}  \notag
\\
&&+\cdots ,  \label{eq:TP1}
\end{eqnarray}%
where the dots denote all effects due to excited and continuum states.

The Eq.\ (\ref{eq:TP1}) can be further detailed using the matrix elements of
the mesons $\eta _{b}$ and $B_{c}^{-}$%
\begin{eqnarray}
\langle 0|J^{\eta _{b}}|\eta _{b}(p^{\prime })\rangle &=&\frac{f_{\eta
_{b}}m_{\eta _{b}}^{2}}{2m_{b}},  \notag \\
\langle 0|J^{B_{c}^{-}}|B_{c}^{-}(q)\rangle &=&\frac{f_{B_{c}}m_{B_{c}}^{2}}{%
m_{b}+m_{c}}.  \label{eq:ME1A}
\end{eqnarray}%
In Eq.\ (\ref{eq:ME1A}) $f_{\eta _{b}}=724~\mathrm{MeV}$ and $%
f_{B_{c}}=(371\pm 37)~\mathrm{MeV}$ are decay constants of the mesons \cite%
{Wang:2024fwc}. Besides, we have to specify the vertex $\langle \eta
_{b}(p^{\prime })B_{c}^{-}(q)|T_{\mathrm{b}}(p)\rangle $, which in the case
of the three spin-$0$ particles has a simple form
\begin{equation}
\langle \eta _{b}(p^{\prime })B_{c}^{-}(q)|T_{\mathrm{b}}(p)\rangle
=G(q^{2})p\cdot p^{\prime }.  \label{eq:SAVAV}
\end{equation}%
Then, we find for $\Pi ^{\mathrm{Phys}}(p,p^{\prime })$
\begin{eqnarray}
&&\Pi ^{\mathrm{Phys}}(p,p^{\prime })=\frac{G(q^{2})\Lambda f_{\eta
_{b}}m_{\eta _{b}}^{2}f_{B_{c}}m_{B_{c}}^{2}}{2m_{b}(m_{b}+m_{c})\left(
p^{2}-m^{2}\right) (p^{\prime 2}-m_{\eta _{b}}^{2})}  \notag \\
&&\times \frac{1}{q^{2}-m_{B_{c}}^{2}}\frac{m^{2}+m_{\eta _{b}}^{2}-q^{2}}{2}%
+\cdots .  \label{eq:PhysS1}
\end{eqnarray}%
The right hand side of this expression is the invariant amplitude $\Pi _{0}^{%
\mathrm{Phys}}(p^{2},p^{\prime 2},q^{2})$ which will be utilized to derive
the sum rule for the form factor $G(q^{2})$. This amplitude can be presented
through double dispersion integral \cite%
{Ioffe:1982qb,Colangelo:2000dp,Agaev:2022iha}

\begin{equation}
\Pi _{0}^{\mathrm{Phys}}(p^{2},p^{\prime 2},q^{2})=\int \int dsds^{\prime }%
\frac{\rho ^{\mathrm{Phys}}(s,s^{\prime },q^{2})}{(s-p^{2})(s^{\prime
}-p^{\prime 2})}+\cdots ,  \label{eq:PhysS1a}
\end{equation}%
where $\rho ^{\mathrm{Phys}}(s,s^{\prime },q^{2})$ is the double spectral
density. The dots in Eq.\ (\ref{eq:PhysS1a}) denote single dispersion
integrals over $s$ and $s^{\prime }$ which vanish after the double Borel
transformation. The $\rho ^{\mathrm{Phys}}(s,s^{\prime },q^{2})$ is given by
the expression
\begin{eqnarray}
&&\rho ^{\mathrm{Phys}}(s,s^{\prime },q^{2})=\widehat{G}(q^{2})\delta
(s-m^{2})\delta (s^{\prime }-m_{\eta _{b}}^{2})  \notag \\
&&+\rho ^{\mathrm{h}}(s,s^{\prime },q^{2})\theta (s-s_{0})\theta (s^{\prime
}-s_{0}^{\prime }).
\end{eqnarray}%
Here, $\widehat{G}(q^{2})$ collects all input constants and functions from
Eq.\ (\ref{eq:PhysS1}), whereas the second term is contributions of the
excited and continuum states encoded by the unknown hadronic spectral
density $\rho ^{\mathrm{h}}(s,s^{\prime },q^{2})$. The unit step functions $%
\theta (s-s_{0})\theta (s^{\prime }-s_{0}^{\prime })$ determine the
boundaries of the a domain $\Sigma $ in the $(s,s^{\prime })$ plane where
these states locate. Then, for $\Pi _{0}^{\mathrm{Phys}}(p^{2},p^{\prime
2},q^{2})$ we get

\begin{eqnarray}
&&\Pi _{0}^{\mathrm{Phys}}(p^{2},p^{\prime 2},q^{2})=\frac{\widehat{G}(q^{2})%
}{\left( p^{2}-m^{2}\right) (p^{\prime 2}-m_{\eta _{b}}^{2})}  \notag \\
&&+\int \int_{\Sigma }dsds^{\prime }\frac{\rho ^{\mathrm{h}}(s,s^{\prime
},q^{2})}{(s-p^{2})(s^{\prime }-p^{\prime 2})}+\cdots .  \label{eq:PhysS2}
\end{eqnarray}

For the QCD side of the sum rule, we obtain
\begin{eqnarray}
&&\Pi ^{\mathrm{OPE}}(p,p^{\prime })=2\int d^{4}xd^{4}ye^{ip^{\prime
}y}e^{-ipx}\left\{ \mathrm{Tr}\left[ \gamma _{5}S_{b}^{ja}(y-x)\right.
\right.  \notag \\
&&\left. \times \gamma _{\mu }\widetilde{S}_{b}^{ib}(-x)\gamma _{5}%
\widetilde{S}_{c}^{ai}(x)\gamma ^{\mu }S_{b}^{bj}(x-y)\right]  \notag \\
&&\left. -\mathrm{Tr}\left[ \gamma _{5}S_{b}^{ja}(y-x)\gamma _{\mu }%
\widetilde{S}_{b}^{ib}(-x)\gamma _{5}\widetilde{S}_{c}^{bi}(x)\gamma ^{\mu
}S_{b}^{aj}(x-y)\right] \right\} .  \notag \\
&&  \label{eq:CF3}
\end{eqnarray}%
The correlator $\Pi ^{\mathrm{OPE}}(p,p^{\prime })$ has also a trivial
Lorentz structure $\sim \mathrm{I}$ and consists of the amplitude $\Pi _{0}^{%
\mathrm{OPE}}(p^{2},p^{\prime 2},q^{2})$. Having applied the double
dispersion relation, we write $\Pi _{0}^{\mathrm{OPE}}(p^{2},p^{\prime
2},q^{2})$ in the form
\begin{equation}
\Pi _{0}^{\mathrm{OPE}}(p^{2},p^{\prime 2},q^{2})=\int_{\mathcal{M}%
^{2}}^{\infty }\int_{4m_{b}^{2}}^{\infty }\frac{dsds^{\prime }\rho ^{\mathrm{%
OPE}}(s,s^{\prime },q^{2})}{(s-p^{2})(s^{\prime }-p^{\prime 2})}+\cdots ,
\label{eq:DDR}
\end{equation}%
where single integrals over $s$ and $s^{\prime }$ arising from subtractions
in the dispersion relation are denoted by the ellipses. In the present work,
the amplitude $\Pi _{0}^{\mathrm{OPE}}(p^{2},p^{\prime 2},q^{2})$ is
calculated by taking into account $\mathrm{Dim4}$ terms $\sim \langle \alpha
_{s}G^{2}/\pi \rangle $. Note that the spectral density $\rho ^{\mathrm{OPE}%
}(s,s^{\prime },q^{2})$ is equal to the imaginary part (a double
discontinuity) of $\Pi _{0}^{\mathrm{OPE}}(s,s^{\prime },q^{2})$.

After equating the functions $\Pi _{0}^{\mathrm{Phys}}(p^{2},p^{\prime
2},q^{2})$ and $\Pi _{0}^{\mathrm{OPE}}(p^{2},p^{\prime 2},q^{2})$,
performing the double Borel transformations over the variables $-p^{2}$, $%
-p^{\prime 2}$ and subtracting from the QCD side of this equality
contributions of the excited and continuum states under the quark-hadron
duality assumption $\rho ^{\mathrm{h}}(s,s^{\prime },q^{2})\simeq \rho ^{%
\mathrm{OPE}}(s,s^{\prime },q^{2})$, we find the sum rule for $G(q^{2})$
\begin{eqnarray}
&&G(q^{2})=\frac{4m_{b}(m_{b}+m_{c})(q^{2}-m_{B_{c}}^{2})}{\Lambda f_{\eta
_{b}}m_{\eta _{b}}^{2}f_{B_{c}}m_{B_{c}}^{2}(m^{2}+m_{\eta _{b}}^{2}-q^{2})}
\notag \\
&&\times e^{m^{2}/M_{1}^{2}}e^{m_{\eta _{b}}^{2}/M_{2}^{2}}\Pi _{0}(\mathbf{M%
}^{2},\mathbf{s}_{0},q^{2}).  \label{eq:SRG}
\end{eqnarray}%
Here $\Pi _{0}(\mathbf{M}^{2},\mathbf{s}_{0},q^{2})$ is given by the
expression
\begin{eqnarray}
\Pi _{0}(\mathbf{M}^{2},\mathbf{s}_{0},q^{2}) &=&\int_{\mathcal{M}%
^{2}}^{s_{0}}\int_{4m_{b}^{2}}^{s_{0}^{\prime }}dsds^{\prime
}e^{-s/M_{1}^{2}-s^{\prime }/M_{2}^{2}}.  \notag \\
&&\times \rho ^{\mathrm{OPE}}(s,s^{\prime },q^{2}).
\end{eqnarray}%
The correlator $\Pi _{0}(\mathbf{M}^{2},\mathbf{s}_{0},q^{2})$ depends on
the parameters $\mathbf{M}^{2}=(M_{1}^{2},M_{2}^{2})$ and $\mathbf{s}%
_{0}=(s_{0},s_{0}^{\prime })$ where the pairs $(M_{1}^{2},s_{0})$ and $%
(M_{2}^{2},s_{0}^{\prime })$ correspond to $T_{\mathrm{b}}$ and $\eta _{b}$
channels, respectively.

For numerical calculations, we should specify $\mathbf{M}^{2}$ and $\mathbf{s%
}_{0}$. Constraints imposed on the auxiliary parameters $\mathbf{M}^{2}$ and
$\mathbf{s}_{0}$ are universal for all SR computations and have been
explained in the previous section. Numerical analysis shows that the regions
in Eq.\ (\ref{eq:Wind1}) for the parameters $(M_{1}^{2},s_{0})$ and
\begin{equation}
M_{2}^{2}\in \lbrack 9,11]~\mathrm{GeV}^{2},\ s_{0}^{\prime }\in \lbrack
95,99]~\mathrm{GeV}^{2}.  \label{eq:Wind3}
\end{equation}%
for $(M_{2}^{2},s_{0}^{\prime })$ satisfy all these requirements. Because,
the form factor $G(q^{2})$ depends on the mass and current coupling of the
tetraquark $T_{\mathrm{b}}$, this choice for $(M_{1}^{2},s_{0})$ excludes
also additional uncertainties in $m$ and $\Lambda $, as well as in $G(q^{2})$
which may appear beyond the regions Eq.\ (\ref{eq:Wind1}).

\begin{figure}[h]
\includegraphics[width=8.5cm]{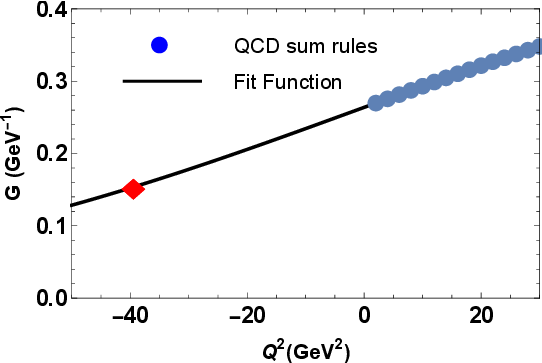}
\caption{The sum rule's data and fit function $\mathcal{G}(Q^{2})$. The red
diamond fixes the point $Q^{2}=-m_{B_c}^{2}$ where $G$ has been estimated. }
\label{fig:Fit}
\end{figure}

The SR method leads to reliable predictions for the form factor $G(q^{2})$
in the Euclidean region $q^{2}<0$. But $G(q^{2})$ determines the strong
coupling $G$ at the mass shell $q^{2}=m_{B_{c}}^{2}$. Therefore, it is
convenient to introduce the function $G(Q^{2})$ with $Q^{2}=-q^{2}$ and use
it in our analysis. The results obtained for $G(Q^{2})$ are plotted in Fig.\ %
\ref{fig:Fit}, where $Q^{2}$ varies inside the limits $Q^{2}=2-30~\mathrm{GeV%
}^{2}$.

As it has been emphasized above, the strong coupling $G$ should be extracted
at $q^{2}=m_{B_{c}}^{2}$, i.e., at $Q^{2}=-m_{B_{c}}^{2}$ where the SR
method does not work. Therefore, we introduce the fit function $\mathcal{G}%
(Q^{2})$ that at momenta $Q^{2}>0$ gives the same SR data, but can be
extrapolated to the domain of negative $Q^{2}$. For these purposes, we
utilize the function
\begin{equation}
\mathcal{G}(Q^{2},m^{2})=\mathcal{G}^{0}\mathrm{\exp }\left[ c^{1}\frac{Q^{2}%
}{m^{2}}+c^{2}\left( \frac{Q^{2}}{m^{2}}\right) ^{2}\right] ,
\label{eq:FitF}
\end{equation}%
where $\mathcal{G}^{0}$, $c^{1}$, and $c^{2}$ are fitted constants. Then,
having compared QCD output and Eq.\ (\ref{eq:FitF}), it is easy to find
\begin{equation}
\mathcal{G}^{0}=0.26~\mathrm{GeV}^{-1},c^{1}=2.75,\text{and }c^{2}=-3.91.
\label{eq:FF1}
\end{equation}%
This function is also shown in Fig.\ \ref{fig:Fit}, where a nice agreement
of $\mathcal{G}(Q^{2})$ and QCD data is clear. For the strong coupling $G$,
we find
\begin{equation}
G\equiv \mathcal{G}(-m_{B_{c}}^{2})=(1.5\pm 0.3)\times 10^{-1}\ \mathrm{GeV}%
^{-1}.  \label{eq:g1}
\end{equation}

The width of the process $T_{\mathrm{b}}\rightarrow \eta _{b}B_{c}^{-}$ is
determined by the expression%
\begin{equation}
\Gamma \left[ T_{\mathrm{b}}\rightarrow \eta _{b}B_{c}^{-}\right] =G^{2}%
\frac{m_{\eta _{b}}^{2}\lambda _{0}}{8\pi }\left( 1+\frac{\lambda _{0}^{2}}{%
m_{\eta _{b}}^{2}}\right) ,  \label{eq:PDw2}
\end{equation}%
where $\lambda _{0}=\lambda (m,m_{\eta _{b}},m_{B_{c}})$
\begin{equation}
\lambda (a,b,c)=\frac{\sqrt{%
a^{4}+b^{4}+c^{4}-2(a^{2}b^{2}+a^{2}c^{2}+b^{2}c^{2})}}{2a}.
\end{equation}

Then, we obtain $\ $%
\begin{equation}
\Gamma \left[ T_{\mathrm{b}}\rightarrow \eta _{b}B_{c}^{-}\right] =(36.0\pm
10.2\mp 1.8)~\mathrm{MeV}.  \label{eq:DW2}
\end{equation}%
The first error above is connected with the uncertainties in the strong
coupling $G$, whereas the second one generated by the errors in the masses
of the mesons $m_{\eta _{b}}$ and $m_{B_{c}}$. We estimate $\Gamma \left[ T_{%
\mathrm{b}}\right] $ as
\begin{equation*}
\Gamma \left[ T_{\mathrm{b}}\right] =(36.0\pm 10.4)~\mathrm{MeV}.
\end{equation*}

\section{Full width of the tetraquark $T_{\mathrm{c}}$}

\label{sec:Widths2}


In this section, we compute the full width of the tetraquark $T_{\mathrm{c}}$
by considering the processes $T_{\mathrm{c}}\rightarrow \eta _{c}B_{c}^{+}$
and $T_{\mathrm{c}}\rightarrow J/\psi B_{c}^{\ast +}$. Both of them are
kinematically allowed decay channels for this particle. In fact, thresholds
for production of the final-state mesons $\eta _{c}B_{c}^{+}$ and $J/\psi
B_{c}^{\ast +}$ are equal to $9259~\mathrm{MeV}$ and $9435~\mathrm{MeV}$
which are smaller that the mass $\widetilde{m}=9680~\mathrm{MeV}$ of the
tetraquark $T_{\mathrm{c}}$.


\subsection{Decay $T_{\mathrm{c}}\rightarrow \protect\eta _{c}B_{c}^{+}$}


Analysis of the process $T_{\mathrm{c}}\rightarrow \eta _{c}B_{c}^{+}$
differ in some technical details from the investigation carried out in the
previous section. Here, we explore the correlation function
\begin{eqnarray}
\widetilde{\Pi }(p,p^{\prime }) &=&i^{2}\int d^{4}xd^{4}ye^{ip^{\prime
}y}e^{-ipx}\langle 0|\mathcal{T}\{J^{B_{c}^{+}}(y)  \notag \\
&&\times J^{\eta _{c}}(0)\widetilde{J}^{\dagger }(x)\}|0\rangle ,
\end{eqnarray}%
where $J^{B_{c}^{+}}(x)$ and $J^{\eta _{c}}(x)$ are the interpolating
currents for the mesons $B_{c}^{+}$ and $\eta _{c}$, respectively:%
\begin{equation}
\ J^{B_{c}^{+}}(x)=\overline{b}_{i}(x)i\gamma _{5}c_{i}(x),\ J^{\eta
_{c}}(x)=\overline{c}_{j}(x)i\gamma _{5}c_{j}(x).
\end{equation}%
The matrix elements required to calculate the phenomenological side of SR
are given the formulas%
\begin{eqnarray}
\langle 0|J^{\eta _{c}}|\eta _{c}(q)\rangle &=&\frac{f_{\eta _{c}}m_{\eta
_{c}}^{2}}{2m_{c}},  \notag \\
\langle 0|J^{B_{c}^{+}}|B_{c}^{+}(p^{\prime })\rangle &=&\frac{%
f_{B_{c}}m_{B_{c}}^{2}}{m_{b}+m_{c}},
\end{eqnarray}%
and
\begin{equation}
\langle \eta _{c}(q)B_{c}^{+}(p^{\prime })|T_{\mathrm{c}}(p)\rangle
=g_{1}(q^{2})p\cdot p^{\prime }.
\end{equation}%
In the expressions above $m_{\eta _{c}}=(2984.1\pm 0.4)~\mathrm{MeV}$ and $%
f_{\eta _{c}}=(421\pm 35)~\mathrm{MeV}$ are the mass and decay constant of
the charmonium $\eta _{c}\ $\cite{PDG:2024,Veliev:2010vd}.

The physical and QCD sides of the sum rule for the form factor $g_{1}(q^{2})$
have the same analytical forms as ones provided in Sec.\ \ref{sec:Widths1}.
Therefore, we write down the SR for $g_{1}(q^{2})$ which reads%
\begin{eqnarray}
&&g_{1}(q^{2}) =\frac{4m_{c}(m_{b}+m_{c})(q^{2}-m_{\eta _{c}}^{2})}{\Lambda
f_{\eta _{c}}m_{\eta
_{c}}^{2}f_{B_{c}}m_{B_{c}}^{2}(m^{2}+m_{B_{c}}^{2}-q^{2})}  \notag \\
&&\times e^{m^{2}/M_{1}^{2}}e^{m_{B_{c}}^{2}/M_{2}^{2}}\Pi _{1}(\mathbf{M}%
^{2},\mathbf{s}_{0},q^{2}),  \label{eq:SRg1}
\end{eqnarray}%
where $\Pi _{1}(\mathbf{M}^{2},\mathbf{s}_{0},q^{2})$ is the
Borel-transformed and subtracted invariant amplitude extracted from the
correlation function $\widetilde{\Pi }^{\mathrm{OPE}}(p,p^{\prime })$.

\begin{figure}[h]
\includegraphics[width=8.5cm]{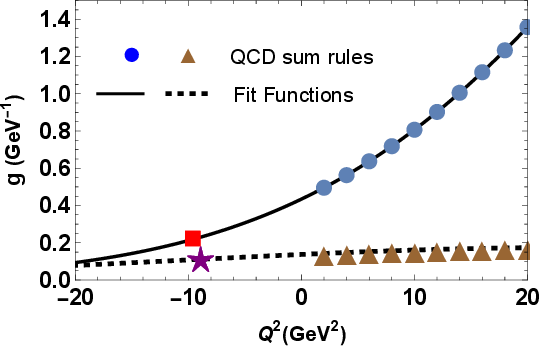}
\caption{Results of the sum rules computations and extrapolating functions $%
\mathcal{F}_{1}(Q^{2})$ (dashed line) and $\mathcal{F}_{2}(Q^{2})$ (solid
line). The red star and square show positions $Q^{2}=-m_{\protect\eta %
_{c}}^2 $ and $Q^{2}=-m_{J/\protect\psi}^2$, respectively. }
\label{fig:Fit1}
\end{figure}

In computations the parameters $(M_{1}^{2},s_{0})$ in the $T_{\mathrm{c}}$
channel are varied within limits Eq.\ (\ref{eq:Wind1A}). The parameters $%
(M_{2}^{2},s_{0}^{\prime })$ in the $B_{c}^{+}$ channel are chosen inside
the regions%
\begin{equation}
M_{2}^{2}\in \lbrack 6.5,7.5]~\mathrm{GeV}^{2},\ s_{0}^{\prime }\in \lbrack
45,47]~\mathrm{GeV}^{2}.  \label{eq:Wind4}
\end{equation}%
The form factor $g_{1}(Q^{2})$ is computed at $Q^{2}=2-20~\mathrm{MeV}^{2}$
(see, Fig.\ \ref{fig:Fit1}). The fit function $\mathcal{F}_{1}(Q^{2},%
\widetilde{m}^{2})$ has the same functional form as the one in Eq.\ (\ref%
{eq:FitF}) but $m^{2}$ replaced by $\widetilde{m}^{2}$. This function in the
case of the form factor $g_{1}(Q^{2})$ has the parameters $\mathcal{F}%
_{1}^{0}=0.14~\mathrm{GeV}^{-1},c_{1}^{1}=1.87,$ and $c_{1}^{2}=-3.58$ and
is depicted in Fig.\ \ref{fig:Fit1}.

The strong coupling $g_{1}$ extracted at the mass shell $q^{2}=-Q^{2}=m_{%
\eta _{c}}^{2}$ amounts to
\begin{equation}
g_{1}\equiv \mathcal{F}_{1}(-m_{\eta _{c}}^{2})=(1.1\pm 0.2)\times 10^{-1}\
\mathrm{GeV}^{-1}.
\end{equation}%
The width of the decay $T_{\mathrm{c}}\rightarrow \eta _{c}B_{c}^{+}$ can be
found by means of the formula
\begin{equation}
\Gamma \left[ T_{\mathrm{c}}\rightarrow \eta _{c}B_{c}^{+}\right] =g_{1}^{2}%
\frac{m_{B_{c}}^{2}\lambda _{1}}{8\pi }\left( 1+\frac{\lambda _{1}^{2}}{%
m_{B_{c}}^{2}}\right) ,
\end{equation}%
with $\lambda _{1}$ being equal to $\lambda (\widetilde{m},m_{B_{c}},m_{\eta
_{c}})$. Our result reads%
\begin{equation}
\Gamma \left[ T_{\mathrm{c}}\rightarrow \eta _{c}B_{c}^{+}\right] =(26.9\pm
6.9\pm 3.5)~\mathrm{MeV}.  \label{eq:DW3}
\end{equation}%
The error $\pm 6.9~\mathrm{MeV}$ in Eq.\ (\ref{eq:DW3}) is due to
ambiguities in the coupling $g_{1}$, while the second error is connected by
uncertainties in the masses $\widetilde{m}$, $m_{B_{c}}$, and $m_{\eta _{c}}$%
.

\subsection{Process $T_{\mathrm{c}}\rightarrow J/\protect\psi B_{c}^{\ast +}$%
}


The second fall-apart mode of the tetraquark $T_{\mathrm{c}}$ is the decay
to the pair of vector mesons $J/\psi B_{c}^{\ast +}$. To determine the
strong coupling $g_{2}$ at the vertex $T_{\mathrm{c}}J/\psi B_{c}^{\ast +}$,
one has to evaluate the corresponding form factor $g_{2}(q^{2})$, which can
extracted from the sum rule for this function. For this purpose, we analyze
the correlation function
\begin{eqnarray}
\Pi _{\mu \nu }(p,p^{\prime }) &=&i^{2}\int d^{4}xd^{4}ye^{ip^{\prime
}y}e^{-ipx}\langle 0|\mathcal{T}\{J_{\mu }^{B_{c}^{\ast }}(y)  \notag \\
&&\times J_{\nu }^{J/\psi }(0)\widetilde{J}^{\dagger }(x)\}|0\rangle ,
\label{eq:CF4}
\end{eqnarray}%
where $J_{\mu }^{B_{c}^{\ast }}(x)$ and $J_{\nu }^{J/\psi }(x)$ are the
interpolating currents of the vector mesons $B_{c}^{\ast +}$ and $J/\psi $.
These current are introduces by means of the expressions%
\begin{equation}
J_{\mu }^{B_{c}^{\ast }}(x)=\overline{b}_{i}(x)\gamma _{\mu }c_{i}(x),\
J_{\nu }^{J/\psi }(x)=\overline{c}_{j}(x)\gamma _{\nu }c_{j}(x).
\end{equation}

The phenomenological side of SR $\Pi _{\mu \nu }^{\mathrm{Phys}}(p,p^{\prime
})$ is given by usual formula%
\begin{eqnarray}
&&\Pi _{\mu \nu }^{\mathrm{Phys}}(p,p^{\prime })=\frac{\langle 0|J_{\mu
}^{B_{c}^{\ast }}|B_{c}^{\ast +}(p^{\prime })\rangle }{p^{\prime
2}-m_{B_{c}^{\ast }}^{2}}\frac{\langle 0|J_{\nu }^{J/\psi }|J/\psi
(q)\rangle }{q^{2}-m_{J/\psi }^{2}}  \notag \\
&&\times \langle B_{c}^{\ast +}(p^{\prime })J/\psi (q)|T_{\mathrm{c}%
}(p)\rangle \frac{\langle T_{\mathrm{c}}(p)|J^{\dagger }|0\rangle }{p^{2}-%
\widetilde{m}^{2}}+\cdots .  \label{eq:CF5}
\end{eqnarray}%
Here, $m_{J/\psi }=(3096.900\pm 0.006)~\mathrm{MeV}$ and $m_{B_{c}^{\ast
}}=6338~\mathrm{MeV}$ are the masses of the mesons $m_{J/\psi }$ and $%
m_{B_{c}^{\ast }}$, respectively. The first of them is the experimental
value of $m_{J/\psi }$ \cite{PDG:2024}, whereas $m_{B_{c}^{\ast }}$-- the
theoretical prediction \cite{Godfrey:2004ya}.

The correlator $\Pi _{\mu \nu }^{\mathrm{Phys}}$ can be simplified by
employing the matrix elements
\begin{eqnarray}
\langle 0|J_{\mu }^{B_{c}^{\ast }}|B_{c}^{\ast +}(p^{\prime })\rangle
&=&f_{B_{c}^{\ast }}m_{B_{c}^{\ast }}\varepsilon _{\mu }^{\prime },  \notag
\\
\langle 0|J_{\nu }^{J/\psi }|J/\psi (q)\rangle &=&f_{J/\psi }m_{J/\psi
}\varepsilon _{\nu },  \label{eq:ME2}
\end{eqnarray}%
and
\begin{eqnarray}
\langle B_{c}^{\ast +}(p^{\prime })J/\psi (q)|T_{\mathrm{c}}(p)\rangle
&=&g_{2}(q^{2})\left[ (q\cdot p^{\prime })(\varepsilon ^{\prime \ast }\cdot
\varepsilon ^{\ast })\right.  \notag \\
&&\left. -(q\cdot \varepsilon ^{\prime \ast })(p^{\prime }\cdot \varepsilon
^{\ast })\right] .
\end{eqnarray}%
In Eq.\ (\ref{eq:ME2}) $f_{J/\psi }=(411\pm 7)~\mathrm{MeV}$ and $%
f_{B_{c}^{\ast }}=471~\mathrm{MeV}$ are the decay constants of the mesons $%
J/\psi $ and $B_{c}^{\ast +}$, respectively \cite%
{Lakhina:2006vg,Eichten:2019gig}.

Then, for $\Pi _{\mu \nu }^{\mathrm{Phys}}(p,p^{\prime })$ we get
\begin{eqnarray}
&&\Pi _{\mu \nu }^{\mathrm{Phys}}(p,p^{\prime })=\frac{g_{2}(q^{2})%
\widetilde{\Lambda }f_{B_{c}^{\ast }}m_{B_{c}^{\ast }}f_{J/\psi }m_{J/\psi }%
}{(p^{2}-\widetilde{m}^{2})(p^{\prime 2}-m_{B_{c}^{\ast
}}^{2})(q^{2}-m_{J/\psi }^{2})}  \notag \\
&&\times \left[ \frac{g_{\mu \nu }(m^{2}-m_{B_{c}^{\ast }}^{2}-q^{2})}{2}%
-q_{\mu }p_{\nu }^{\prime }+\text{other~structures}\right] .  \notag \\
&&  \label{eq:CF5a}
\end{eqnarray}%
The correlation function $\Pi _{\mu \nu }(p,p^{\prime })$ expressed in terms
of the heavy quark propagators is%
\begin{eqnarray}
&&\Pi _{\mu \nu }^{\mathrm{OPE}}(p,p^{\prime })=2\int
d^{4}xd^{4}ye^{ip^{\prime }y}e^{-ipx}\left\{ \mathrm{Tr}\left[ \gamma _{\nu
}S_{c}^{ja}(-x)\right. \right.  \notag \\
&&\left. \times \gamma _{\alpha }\widetilde{S}_{c}^{ib}(y-x)\gamma _{\mu }%
\widetilde{S}_{b}^{ai}(x-y)\gamma ^{\alpha }S_{c}^{bj}(x)\right]  \notag \\
&&\left. -\mathrm{Tr}\left[ \gamma _{\nu }S_{b}^{ja}(-x)\gamma _{\alpha }%
\widetilde{S}_{c}^{ib}(y-x)\gamma _{\mu }\widetilde{S}_{b}^{bi}(x-y)\gamma
^{\alpha }S_{c}^{aj}(x)\right] \right\} .  \notag \\
&&  \label{eq:CF6}
\end{eqnarray}%
To derive SR for the form factor $g_{2}(q^{2})$, we utilize the invariant
amplitudes $\Pi _{2}^{\mathrm{Phys}}(p^{2},p^{\prime 2},q^{2})$ and $\Pi
_{2}^{\mathrm{OPE}}(p^{2},p^{\prime 2},q^{2})$ which correspond to terms
proportional to $-q_{\mu }p_{\nu }^{\prime }$ in Eqs.\ (\ref{eq:CF5a}) and (%
\ref{eq:CF6}). As a result, we find for $g_{2}(q^{2})$%
\begin{eqnarray}
&&g_{2}(q^{2})=\frac{(q^{2}-m_{J/\psi }^{2})}{\widetilde{\Lambda }%
f_{B_{c}^{\ast }}m_{B_{c}^{\ast }}f_{J/\psi }m_{J/\psi }}  \notag \\
&&\times e^{m^{2}/M_{1}^{2}}e^{m_{B_{c}^{\ast }}^{2}/M_{2}^{2}}\Pi _{2}(%
\mathbf{M}^{2},\mathbf{s}_{0},q^{2}).
\end{eqnarray}

Operations to calculate the strong coupling $g_{2}$ of particles at the
vertex $T_{\mathrm{c}}J/\psi B_{c}^{\ast +}$ have been described above,
therefore, we provide relevant information in a brief form. Thus, the form
factor $g_{2}(Q^{2})$ is computed at $Q^{2}=2-20~\mathrm{MeV}^{2}$. In
calculations the parameters $(M_{1}^{2},s_{0})$ in the $T_{\mathrm{c}}$
channel are chosen in accordance with Eq.\ (\ref{eq:Wind1A}). In the $%
B_{c}^{\ast +}$ channel the parameters $(M_{2}^{2},s_{0}^{\prime })$ change
in the regions
\begin{equation}
M_{2}^{2}\in \lbrack 6.5,7.5]~\mathrm{GeV}^{2},\ s_{0}^{\prime }\in \lbrack
50,51]~\mathrm{GeV}^{2}.
\end{equation}%
The extrapolating function $\mathcal{F}_{2}(Q^{2},\widetilde{m}^{2})$ has
the following parameters: $\mathcal{F}_{2}^{0}=0.43~\mathrm{GeV}%
^{-1},c_{2}^{1}=6.26,$and $c_{2}^{2}=-4.33$. Then, the coupling $g_{2}$ is
equal to
\begin{equation}
g_{2}\equiv \mathcal{F}_{2}(-m_{J/\psi }^{2})=(2.2\pm 0.4)\times 10^{-1}\
\mathrm{GeV}^{-1}.
\end{equation}

The width of the decay $T_{\mathrm{c}}\rightarrow J/\psi B_{c}^{\ast +}$ is
determined by employing the expression
\begin{equation}
\Gamma \left[ T_{\mathrm{c}}\rightarrow J/\psi B_{c}^{\ast +}\right]
=g_{2}^{2}\frac{\lambda _{2}^{3}}{4\pi }\left( 1+\frac{3m_{B_{c}^{\ast
}}^{2}m_{J/\psi }^{2}}{2\widetilde{m}^{2}\lambda _{2}^{2}}\right) ,
\end{equation}%
where $\lambda _{2}$ is $\lambda (\widetilde{m},m_{B_{c}^{\ast }},m_{J/\psi
})$. We find%
\begin{equation}
\Gamma \left[ T_{\mathrm{c}}\rightarrow J/\psi B_{c}^{\ast +}\right]
=(27.8\pm 7.1\pm 6.9)~\mathrm{MeV}.
\end{equation}%
The full width of the tetraquark $T_{\mathrm{c}}$ saturated by these two
decay channels is%
\begin{equation}
\Gamma \left[ T_{\mathrm{c}}\right] =(54.7\pm 12.6)~\mathrm{MeV}.
\end{equation}


\section{Conclusions}

\label{sec:Conc}


In the current paper, we have calculated the masses and widths of the scalar
tetraquarks $T_{\mathrm{b}}=bb\overline{b}\overline{c}$ and $T_{\mathrm{c}%
}=cc\overline{c}\overline{b}$ in the context of QCD sum rule method. These
particles have been modeled as diquark-antidiquark systems made of the
axial-vector diquarks and antidiquarks. Their masses have been evaluated
using the two-point sum rule approach, whereas to estimate the widths of
these tetraquarks we have invoking techniques of the three-point sum rule
approach. Here, we have considered only the dissociations of $T_{\mathrm{b}}$
and $T_{\mathrm{c}}$ to conventional mesons, which are their dominant decay
channels. There are also modes generated by annihilations of $b\overline{b}$
and $c\overline{c}$ quarks inside of $T_{\mathrm{b}}$ and $T_{\mathrm{c}}$.
But, the partial widths of these processes, as usual, are smaller than that
of the dominant channels and have not been taken into account in this
article.

The tetraquarks $T_{\mathrm{b}}$ and $T_{\mathrm{c}}$ have different
parameters and features. Thus, $T_{\mathrm{b}}$ and $T_{\mathrm{c}}$ bear a
unit of negative and positive electric charge, respectively. The gap between
the masses of these particles is approximately equal to the mass difference
of corresponding constituent quarks. The structure $T_{\mathrm{b}}$ with the
width $36.0~\mathrm{MeV}$ is narrower than the tetraquark $T_{\mathrm{c}}$.
The reason is that the mass $m$ of $T_{\mathrm{b}}$ exceeds only $\eta
_{b}B_{c}^{-}$ threshold, whereas $\widetilde{m}$ overshoots kinematical
limits for productions of $\eta _{c}B_{c}^{+}$ and $J/\psi B_{c}^{\ast +}$
mesons.

We have noted that calculation of the $T_{\mathrm{b}}$ and $T_{\mathrm{c}}$
tetraquarks' masses in the framework of quark models led to different
predictions. Based on these results, in some articles, the authors made
assumptions about widths of these particles. Thus, in Ref.\ \cite%
{Liu:2019zuc} the mass of the tetraquark $T_{\mathrm{c}}$ was found equal to
$9740~\mathrm{MeV}$ (color-triplet diquarks) and $9763~\mathrm{MeV}$
(color-sextet diquarks). This is $300-320~\mathrm{MeV}$ above the $J/\psi
B_{c}^{\ast +}$ threshold, therefore in experiments one has to observe broad
$cc\overline{c}\overline{b}$ states. The mass of the tetraquark $T_{\mathrm{c%
}}$ obtained in the present article exceeds the $J/\psi B_{c}^{\ast +}$
threshold about $245~\mathrm{MeV}$, but width of the decay $T_{\mathrm{c}%
}\rightarrow J/\psi B_{c}^{\ast +}$ is equal to approximately $30~\mathrm{MeV%
}$ and is not large for tetraquarks. It should be noted that the width of
the process $T_{\mathrm{c}}\rightarrow J/\psi B_{c}^{\ast +}$ depends not
only on the available large phase-space, but also on the strong coupling $%
g_{2}$ at the vertex $T_{\mathrm{c}}J/\psi B_{c}^{\ast +}$ and expression
for the width under consideration. Our computations show that both the
tetraquarks $T_{\mathrm{b}}$ and $T_{\mathrm{c}}$ are relatively narrow
states.

In Sec.\ \ref{sec:Mass} it has been pointed out that the scalar particles $%
T_{\mathrm{b}}$ and $T_{\mathrm{c}}$ may have alternative
diquark-antidiquark organizations. Due to differences in the internal
structures, parameters of such states would differ from those of $T_{\mathrm{%
b}}$ and $T_{\mathrm{c}}$. The tetraquarks $bb\overline{b}\overline{c}$ and $%
cc\overline{c}\overline{b}$ are still hypothetical particles and were not
observed in experiments. The physical resonances may have various
organizations: First, they may demonstrate features of pure $C\gamma _{\mu
}\otimes \gamma ^{\mu }C$ and/or $C\sigma _{\mu \nu }\otimes \sigma _{\mu
\nu }C$ states. Alternatively, measured parameters may reveal their mixed
nature, i.e., the physical resonances may appear in future experiments as
superpositions of pure diquark-antidiquark states. Any mixing scheme is
characterized by basic states and a mixing angle that should be extracted
from comparison of theoretical predictions and experimental data.   Analysis performed in the present work, in the absence of such data, provides valuable information about scalar tetraquarks $bb\overline{b}\overline{c}$ and $cc\overline{c}\overline{b}$. However, it is important to reinforce, as mentioned above, that the whole picture could correspond to the present results, or to those obtained by the second current, or to those corresponding to a mixture of the two types of currents. Such a complete study is beyond the scope of the present work.

In our previous articles, we performed rather detailed investigations of the
fully heavy tetraquarks. With related information at hand, it is instructive
to compare the masses of the fully heavy scalar tetraquarks containing
different number of valence $b(\overline{b})$ quarks and analyze their
stability against dissociation to ordinary mesons. The four-quark meson
containing one $\overline{b}$ quark has been explored in the present paper.
Its mass exceeds $421~\mathrm{MeV}$ the lowest $\eta _{c}B_{c}^{+}$
threshold which makes possible decays to $\eta _{c}B_{c}^{+}$ and $J/\psi
B_{c}^{\ast +}$ pairs.

There are two classes of particles with two constituent $b(\overline{b})$
quarks. The first class includes states with $bb$ diquarks or $\overline{b}%
\overline{b}$ antidiquarks, whereas the hidden charm-bottom tetraquarks form
the second type of such particles: They have the $bc\overline{b}\overline{c}$
structures. The scalar exotic meson $bb\overline{c}\overline{c}$ was
explored in our paper \cite{Agaev:2023tzi}, in which its mass was found
equal to $12715~\mathrm{MeV}$, i.e., approximately $166~\mathrm{MeV}$ above
the $2B_{c}^{-}$ mesons' mass. The scalar tetraquark $bc\overline{b}%
\overline{c}$ with the mass $12697~\mathrm{MeV}$ is $166~\mathrm{MeV}$
heavier than the lowest $B_{c}^{-}B_{c}^{+}$ final-state \cite{Agaev:2024wvp}.

The scalar diquark-antidiquark state $T_{\mathrm{b}}=bb\overline{b}\overline{%
c}$ have been explored in the present work and found only $25~\mathrm{MeV}$
exceeding the $\eta _{b}B_{c}^{-}$ mass limit. Finally, the scalar
tetraquark $X_{4\mathrm{b}}=bb\overline{b}\overline{b}$ is below the $\eta
_{b}\eta _{b} $ threshold and stable against decays to these mesons \cite%
{Agaev:2023wua}. But $X_{4\mathrm{b}}$ can still transform to ordinary
particles through strong interaction due to annihilation of $b\overline{b}$
quarks and production of $B_{q}\overline{B}_{q}$ and $B_{q}^{\ast }\overline{%
B}_{q}^{\ast }$ pairs \cite{Agaev:2023ara}.

As is seen, stability of the fully heavy tetraquarks against fall-apart
processes and their widths depend on a number of valence $b(\overline{b})$
quarks. Our analysis of all-heavy exotic mesons and collected theoretical
information on their masses and widths can be used in analyses of future
experimental data.

\section*{ACKNOWLEDGEMENTS}

K. Azizi thanks Iran national science foundation (INSF) for the partial
financial support provided under the elites Grant No. 4037888. He is also
grateful to the CERN-TH department for their support and warm hospitality.

\end{document}